\documentclass[aps,prc,twocolumn,superscriptaddress]{revtex4-1}

\usepackage{graphics}
\usepackage{amsmath}
\usepackage[normalem]{ulem}
\usepackage{comment}
\usepackage{epsfig}
\usepackage{color}  % only for remarks
\usepackage{hyperref}
\usepackage{soul}

\def\nuc#1#2{\relax\ifmmode{}^{#1}{\protect\text{#2}}\else${}^{#1}$#2\fi}

\newcommand{\be}{\begin{eqnarray}}
\newcommand{\ee}{\end{eqnarray}}

\begin{document}

%\preprint{}

\title{Evidence of strong dynamic core excitation in  $^{19}$C resonant break-up} 


%\author{}

\author{J.~A. Lay}
\email{lay@pd.infn.it}
\affiliation{Dipartimento di Fisica e Astronomia, Universit\`{a} di Padova, I-35131 Padova, Italy }
\affiliation{Istituto Nazionale di Fisica Nucleare, Sezione di Padova, I-35131 Padova, Italy}

\author{R.~de Diego}
%\email{raulez@cii.fc.ul.pt}
%\altaffiliation{Present address: Centro de F\'{\i}sica Nuclear,
%Universidade de Lisboa, Av. Prof. Gama Pinto 2, P-1649-003 Lisboa,
%Portugal}% 
\affiliation{Centro de Ci\^{e}ncias e Tecnologias Nucleares, Universidade de Lisboa, Estrada Nacional 10 (Km 139,7), P-2695-066 Loures, Portugal.}%

\author{R.~Crespo}
%\altaffiliation{Present address: Centro de F\'{\i}sica Nuclear,
%Universidade de Lisboa, Av. Prof. Gama Pinto 2, P-1649-003 Lisboa,
%Portugal}% 
\affiliation{Centro de Ci\^{e}ncias e Tecnologias Nucleares, Universidade de Lisboa, Estrada Nacional 10 (Km 139,7), P-2695-066 Loures, Portugal.}%

\author{A.~M.\ Moro}
%\email{moro@us.es}
\affiliation{Departamento de FAMN, Facultad de F\'{\i}sica, Universidad de Sevilla, Apdo.~1065, E-41080 Sevilla, Spain}

\author{J.~M.\ Arias}
%\email{ariasc@us.es}
\affiliation{Departamento de FAMN, Facultad de F\'{\i}sica, Universidad de Sevilla, Apdo.~1065, E-41080 Sevilla, Spain}

\author{R.~C.\ Johnson}
\affiliation{Physics Department, University of Surrey, Guildford, Surrey GU2 7XH, U. K.}

%\vspace{1cm}

\date{\today}

%\thanks{}
%\affiliation{}

\begin{abstract} 

The resonant break-up of $^{19}$C on protons measured at RIKEN [Phys. Lett. B 660, 320 (2008)] is analyzed in terms of a valence-\textit{core} model for $^{19}$C including possible \textit{core} excitations. The analysis of the angular distribution of a prominent peak appearing in the relative-energy spectrum could be well described with this model and is consistent with the previous assignment of $5/2^{+}$ for this state. Inclusion of core-excitation effects are found to be essential to give the correct magnitude of the cross section for this state.  By contrast, the calculation assuming an inert $^{18}$C core is found to largely underestimate the data. 
\end{abstract}

%\pacs{21.10.Jx, 21.60.-n 24.10.Eq, 27.20.+n}
%\pacs{ 24.10.Eq, 25.10.+s, 25.45.De}
%\keywords{}
%24.10.-i,, 25.60.Gc
\maketitle

\textit{Introduction.} Current developments in radioactive beam facilities are permitting the production of neutron-rich nuclei which are both farther away from the stability line and heavier in mass. Among them, exotic structures, such as \textit{haloes}, continue to receive special attention due to their remarkable properties.  These nuclei are characterized by the presence of one or two weakly bound nucleons, which can thereby explore distances far from the rest of the nucleus, usually referred to as \textit{core}. This decoupling of the valence particle(s) with respect to the tighter core permits to study the structure and reactions of these systems in terms of few-body models.

In reactions involving halo nuclei, break-up channels  are enhanced due to their small binding energy. In the case of elastic breakup, the standard formalisms to study these reactions are the continuum-discretized coupled-channels (CDCC) method~\cite{Ron70,Raw74,Aus87,Tho09}, the adiabatic approximation~\cite{Tos98,Ban00} and different semiclassical approximations~\cite{Typ94,Esb96}. Recently, it has become possible to solve AGS-Faddeev equations for specific cases \cite{Alt,Del09}. 

In their standard formulations, the target and the constituent fragments of the projectile are considered to be inert and, therefore, possible excitations of them are ignored. The assumption of inert fragments is well justified for reactions with deuterons, where these formalisms were first applied~\cite{Ron70}. It is expected to be a good approximation for the traditional two-neutron halo nuclei $^{6}$He and $^{11}$Li. However, in odd nuclei with a well deformed core, such as in the $^{11}$Be or $^{19}$C cases, the inert-core approximation is less justified. For $^{11}$Be, the archetype of one-neutron halo nucleus, the single-particle picture based on a neutron orbiting a $^{10}$Be(g.s.) core provides a rough description of the low-lying spectrum of this nucleus. The model  has also permitted a reasonable description of nuclear reactions, assuming that the contributions of core-excited admixtures can be included in an effective way. For example, in transfer reactions this is usually done multiplying the inert-core result by the corresponding spectroscopic factor. Dynamic core excitations (DCE) occurring during the collision are effectively included in the effective core-target  potentials. 

However, there is evidence that this approximate model is not always accurate~\cite{Cre11,Mor12a,Mor12}. For example, recent calculations \cite{Che16} have shown that, in collisions of $^{11}$Be with light targets, the explicit inclusion of the DCE mechanism gives rise to a sizable increase of the breakup cross section. This is particularly important for excitation energies around the low-lying $3/2^+$ resonance, where the effect is enhanced due to the dominant $^{10}$Be(2$^+$)  configuration for this resonance. Moreover, the admixtures of different core states in the $^{11}$Be states modifies the shape of the breakup angular distribution \cite{Mor12a}. 
%it should be noted that in situations where the first $3/2^{+}$ resonance in $^{11}$Be plays an important role, since it is mainly based on the $^{10}$Be first excited state, one is entitled to include explicitly the core excitations. Anyway, the minimum excitation energy in $^{10}$Be is still relatively high. The situation will increase in complexity for heavier neutron-rich halo nuclei since the core excitation energies will be smaller.
We expect that these effects will show up in other deformed weakly-bound nuclei.
%, particularly in those cases where the excitation energy of the core is small, thereby favoring the excitation mechanism.  
This is the case of $^{19}$C,  where the core, $^{18}$C, is well deformed and has a  first excited 2$^{+}$state at $1.6$~MeV. In addition, new halo candidates like $^{31}$Ne and $^{37}$Mg are within a well-established deformed region. % for which relatively small core excitation energies are expected and, consequently, they will not fit the image of an inert-core-plus-valence particle.
Therefore, deviations from the naive inert-core-plus-valence particle are expected. We note that these dynamic core excitation effects have been also recently studied in the context of transfer reactions~\cite{Gom15,Del13}.

The success of few-body models describing halo phenomena suggests that the presence of a halo always implies a decoupling of its motion from the excitations of the core. As we mention here, there are several cases in the literature where the inclusion of core excitations and their interplay with excitations of the valence particle are mandatory to understand the experimental data~\cite{Mor12a,Love67}. The novelty in the case we are discussing here, $^{19}$C, is that the resonant break-up cross section is dominated almost entirely by the dynamic excitation of the core. This is so strong that is able to overwhelm the role of the halo as we will demonstrate in the following. 

Despite the increased complexity, the study of core excitations constitutes a great opportunity to deepen our knowledge on these new structures. For example, it was shown in~\cite{Mor12a} that the presence of different core states  admixtures has a sizable impact in the resonant break-up of halo nuclei. By analyzing these reactions one can extract information on the relative weights of the different core states in the spectra of the halo nucleus of interest.  It also provides spectroscopic information on resonances which are weakly populated in transfer reactions, that is a more standard spectroscopic probe.

In the last years, some of the traditional formalisms for studying break-up have been upgraded to include static and dynamic excitations of the core during the reaction process. This is the case of the no-recoil XDWBA~\cite{Mor12}, the XCDCC method~\cite{Summers06,*Sum14,deD14}, and a new formulation of the AGS-Faddeev equations~\cite{Del13,Del15}. Most of them focus on $^{11}$Be as a benchmark. Here we focus on the less-known case of $^{19}$C.

\begin{figure}
\epsfig{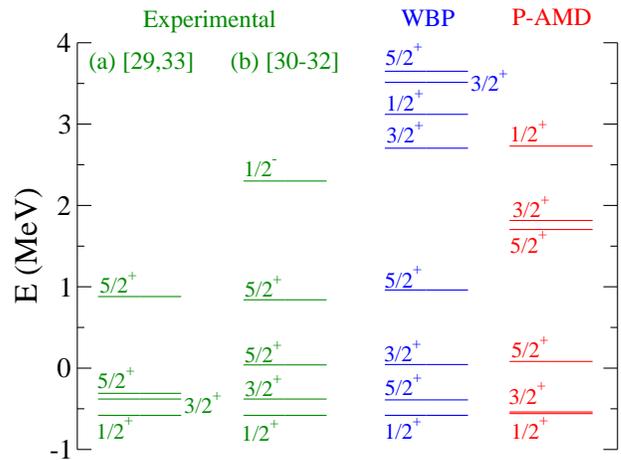}
\caption{ (Color online) Experimental (far left column)~\cite{Ele05,Sat08} and calculated spectrum of $^{19}$C in a shell-model calculation with \texttt{OXBASH} using the WBP interaction~\cite{OXB,WBT} (middle right column), and within a semi-microscopic core-plus-valence-particle calculation (P-AMD, far right column) which takes into account core excitations as described in~\cite{Lay14}. We include a second experimental spectrum (middle left column) due to the discrepancy raised by  the latest knock-out experiments~\cite{Kob12,Vaj15,Hwang16}.}
\label{c19levels}
\end{figure}

The $^{19}$C nucleus has raised interest in connection with the disappearance of the N=14 shell closure and the emergence of a subshell closure at N=16~\cite{Ele09} and the possible shape-phase transition from prolate to oblate in the carbon isotopic chain~\cite{Suz03,Sag04}. $^{19}$C is a halo nucleus~\cite{Oza01b} with a well stablished spin 1/2$^{+}$ ground state~\cite{Baz95,Bau98,Nak99} and a neutron separation energy $\epsilon_B=$ 0.589~MeV~\cite{Aud03}. The situation is controversial for the rest of the low-lying spectrum. Two bound states, $3/2^{+}$ and $5/2^{+}$, with respect to neutron emission were proposed in Ref.~\cite{Ele05} (see left column in Fig.~\ref{c19levels}). Although this is supported by $sd$ shell-model calculations  (middle right column of Fig.~\ref{c19levels}), the existence of a bound $5/2^{+}$ state seems to be excluded according to knock-out experiments~\cite{Kob12,Vaj15,Hwang16}. In addition, an unbound $5/2^{+}$ state was found at RIKEN in the break-up of $^{19}$C on protons at 70 MeV/nucleon~\cite{Sat08} and more recently in a one-neutron knockout reaction at 290 MeV/nucleon~\cite{Hwang16}. Semi-microscopic predictions and shell-model calculations suggest a strong overlap of this state with the 2$^{+}$ core excited state~\cite{Lay14}. Nevertheless, both of them, and even ab-initio coupled-cluster calculations, produce two $5/2^{+}$ states within the first 2~MeV of excitation energy.

The resonant break-up cross section found in Ref.~\cite{Sat08} and associated to a $5/2^{+}$ state was previously analyzed in~\cite{Cre11b} within an inert-core AGS-Faddeev formalism using a realistic CD-Bonn interaction. Single particle excitation was unable to explain the data by an order of magnitude, thus being a motivation to explore the role of core excitations in this nucleus.

In the following, we will analyze the data from Ref.~\cite{Sat08} in order to clarify the nature of the measured $5/2^{+}$ unbound state. Following \cite{Lay14,Lay12}, we describe the $^{19}$C nucleus using a core-plus-valence-particle model, including core excitations,  and will compute the resonant break-up using the extended versions of the DWBA~\cite{Mor12,Mor12a} and the CDCC formalisms~\cite{Summers06,*Sum14,deD14}. Through this analysis we will show how in this reaction the core excitation role is by far dominant. The importance of core excitations is much larger than in the previously analyzed case, $^{11}$Be, due to the $^{18}$C lower excitation energy and its larger deformation. Similar effects might be expected for more exotic halo candidates like the aforementioned $^{31}$Ne and $^{37}$Mg.
 
   %This larger impact than in the previously analyzed case, $^{11}$Be, can be understood thanks to the $^{18}$C lower excitation energy and its larger deformation. 

%-------------------------------------------
\textit{Structure and reaction formalisms.} 
%--------------------------------------------
Further details of the core excitation model used in this work can be found in Refs.~\cite{deD14,Mor12,Mor12a,Lay12}. Here, only the main ingredients are briefly discussed. We consider the reaction of a two-body weakly-bound projectile ($^{19}$C  in our case) on a proton target. We describe the projectile in the weak-coupling limit, using a core+valence-particle model ($^{18}$C+$n$). A general projectile wavefunction for this model can be expanded as:
\begin{equation}
\Psi_{JM}(\vec{r},\xi)=\sum_{\alpha}\left[\varphi_{\alpha}(\vec{r})\otimes\Phi_{I}(\xi)\right]_{JM},
\end{equation}
where the functions $\varphi_{\alpha}(\vec{r})$ describe the relative motion between the valence particle and the core, and $\Phi_{IM_{c}}(\xi)$ are the core eigenstates with angular momentum $I$ and projection $M_{c}$. $\xi$ represents the core internal degrees of freedom. The index $\alpha$ denotes the set of quantum numbers $\{l,s,j,I\}$, with $l$, $s$, and $j$ being the orbital angular momentum, the intrinsic spin of the valence particle, and their sum ($\vec{j}=\vec{l}+\vec{s}$), respectively. Any wavefunction will be sum of different configurations (channels) labeled here with the parameter $\alpha$. Each channel will have a specific weight in each state of the composite nucleus. This weight can be regarded as a unit-normalized spectroscopic factor. 

Once defined the structure model, for the reaction calculations one needs also the optical potential representing the interaction of the projectile with the target. Within the assumed three-body reaction model, this interaction will be the sum of the interactions of the different projectile constituents (core + valence) with the target ($T$), i.e.:
\begin{equation}
V_{pT}=V_{vT}(\vec{R_{vT}})+V_{cT}(\vec{R_{cT}},\xi).
\end{equation}
$V_{vT}$ and $V_{cT}$ are evaluated at the energy per nucleon of the incident projectile.
This interaction enters in the reaction calculation through the coupling potentials or form factors which read:
\begin{equation}
%\left\langle \chi_{\vec{K_{i}}}(\vec{R})\Psi^{i}_{JM} \left| V_{vT}(\vec{R_{vT}})+V_{cT}(\vec{R_{cT}},\xi) \right|\chi_{\vec{K_{f}}}(\vec{R})\Psi^{f}_{J'M'} \right\rangle.
\left\langle  \Psi^{f}_{JM} \left| V_{vT}(\vec{R_{vT}})+V_{cT}(\vec{R_{cT}},\xi) \right| \Psi^{i}_{J'M'} \right\rangle.
\end{equation}
Note that this $V_{cT}$ depends, in addition to the relative coordinates, on the  core internal degrees of freedom, $\xi$. In this way, the core-target interaction is able to excite the core states during the reaction process. This implies to connect and explore wavefunction parts not accessible through the \textit{normal} valence particle excitation, which is the aspect we intend to exploit with this kind of analysis. This process is normally called {\it dynamic} core excitation (DCE) to distinguish it from the {\it static} effect of these excitations in the projectile structure. In other words, {\it static} effects are connected to the weights of the different contributions in the wavefunctions of the projectile $\Psi_{JM}$, whereas {\it dynamic} effects are related to $V_{cT}$. Standard few-body models neglect this dependence of $V_{cT}$ on $\xi$, thereby omitting the dynamical excitation of the core.

We will use here two different frameworks which are the appropriate generalizations of the DWBA and CDCC formalisms for break-up reactions including both static and dynamical core excitations. 
%The main ingredients in both cases are the coupling potentials or form factors which read:
%%
%\begin{equation}
%%\left\langle \chi_{\vec{K_{i}}}(\vec{R})\Psi^{i}_{JM} \left| V_{vT}(\vec{R_{vT}})+V_{cT}(\vec{R_{cT}},\xi) \right|\chi_{\vec{K_{f}}}(\vec{R})\Psi^{f}_{J'M'} \right\rangle.
%\left\langle  \Psi^{i}_{JM} \left| V_{vT}(\vec{R_{vT}})+V_{cT}(\vec{R_{cT}},\xi) \right| \Psi^{f}_{J'M'} \right\rangle.
%\end{equation}
% $\chi$ represents the part of the total wavefunction that corresponds to the relative target-projectile motion. In DWBA calculations, they are replaced by distorted waves.
The main difference between the XDWBA and XCDCC approaches is that, in the former, the breakup is treated to first order and the relative motion of the projectile and target is described by appropriate distorted waves, whereas in the CDCC formalism the breakup is treated to all orders, and the functions describing the projectile-target relative motion are obtained by solving a system of coupled equations. Additionally, in the XDWBA method used here, we make a {\it no-recoil approximation}, in which the core-target coordinate is approximated by the projectile-target coordinate. These two approximations are expected to be well justified in the present case \cite{Mor12c}. 
%Further details can be found in \cite{Mor12}
%The no-recoil XDWBA can be understood as a first order version of the full XCDCC calculation. In reactions at intermediate energies, like the one we are considering here, both calculations are expected to give similar results~\cite{}.
In addition to simplifying the reaction problem, the appealing feature of the XDWBA formalism is that it permits a separation of the scattering amplitude into two terms: one corresponding to the excitation of the valence particle and the other one associated with the core excitation. We will take advantage of this separation to evaluate the relative importance of the two processes: i) the traditional elastic break-up due to the excitation of the weakly-bound neutron and ii) the break-up due to the dynamical core excitation where the valence neutron is just a spectator.

\textit{Results.} We apply the XDWBA and  XCDCC frameworks to the resonant break-up of $^{19}$C on protons at 70 MeV/nucleon. This reaction was measured at RIKEN by Satou {\it et al.}~\cite{Sat08}. In this experiment they found a prominent peak in the energy distribution of the break-up cross section at $E_{x}=1.46\pm0.10$~MeV. From a microscopic DWBA analysis of the corresponding angular distribution, this peak was associated with a resonance with spin and parity 5/2$^+$. However, different structure models predict two 5/2$^+$ resonances and there is a long standing controversy regarding the possibility of having a 5/2$^+$ bound state, as suggested by Elekes {\it et al.}~\cite{Ele05}.

In the present calculations, we will consider the recently developed semi-microscopic particle-core model for $^{19}$C \cite{Lay14}, in which the diagonal and off-diagonal neutron-core couplings are obtained by folding the effective JLM interaction \cite{JLM} with microscopic central and transition densities of $^{18}$C, calculated with 
Antisymmetrized Molecular Dynamics (AMD) \cite{Kan13b}. For simplicity, only the $0^+$ and $2^{+}$ states of the core are considered, and the orbital angular momentum of the halo neutron is restricted to $\ell=0,2$. A phenomenological spin-orbit potential with standard parameters is also added. The wavefunctions and energies of the system are then obtained by diagonalizing this Hamiltonian in a transformed harmonic oscillator (THO) basis. With a suitable choice of the basis, the resonance states are well characterized by a single eigenstate.  Further details can be found in Ref.~\cite{Lay14}. The resulting low-lying spectrum is depicted in the last column of Fig.~\ref{c19levels}. Despite its simplicity, the model succeeds in reproducing the doublet of bound states $1/2^+$ and $3/2^+$. It predicts two unbound  $5/2^+$ resonances. This is in disagreement with the observations of Ref.~\cite{Ele05}, but is consistent with the observation of Ref.~\cite{Vaj15} and also with the conclusions of Ref.~\cite{Kob12}. However, none of these two states has an energy consistent with the peak observed by Satou {\it et al.}~\cite{Sat08}. Consequently, it is not possible to assign the peak observed by Satou to one of our $5/2^+$ states based solely on their energies. Thus,  in the reaction calculations we have considered both resonances as potential candidates for this peak.

For both the XDWBA and XCDCC calculations, valence-target and core-target interactions are also needed. For the $p$-$^{18}$C interaction we construct folding potentials using the JLM nucleon-nucleon interaction \cite{JLM}. This procedure has been able to reproduce the elastic and inelastic scattering of protons on $^{10}$Be and $^{12}$Be \cite{Tak08}, after some suitable renormalization of the real and imaginary parts. The renormalization factors depend also on the assumed range parameter for the JLM interaction ($t$). We adopt here the original value, $t=1.4$~fm, for which renormalization factors of $1.2$ and  $0.8$ have been prescribed for the real and imaginary parts, respectively.

For the $n$-$p$ potential, we use the simple Gaussian potential of Refs.~\cite{Cre11,Mor12}, whose parameters were adjusted to reproduce the breakup in the  $^{11}$Be+p  
 reaction obtained with a Faddeev calculation with the more realistic $p$-$n$ CD-Bonn potential.

\begin{figure}

\epsfig{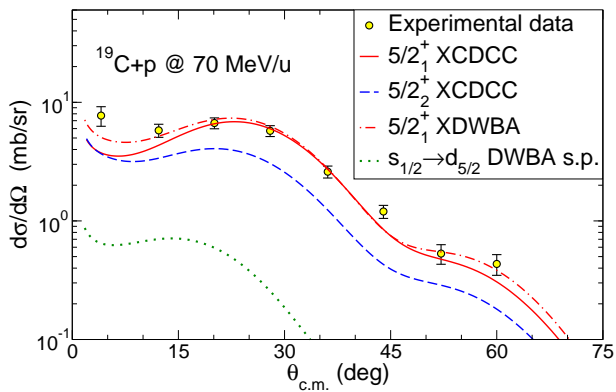}
\caption{ (Color online) Angular distribution of the resonant break-up of $^{19}$C on protons at $70$~MeV/u. The solid red line and the dashed blue line correspond to the XCDCC calculation for the first and the second 5/2$^{+}$ resonance of the P-AMD model~\cite{Lay14} respectively. The dotted dashed line corresponds to a XDWBA calculation for the first 5/2$^{+}$ resonance. 
The dotted line corresponds to an inert-core DWBA calculation where ground state and resonance are considered to be pure $s_{1/2}$ and $d_{5/2}$ states respectively. Experimental data is from Ref.~\cite{Sat08}.}
\label{xsecconv}
\end{figure}

The calculated break-up angular distribution for the two 5/2$^{+}$ resonances predicted by our structure model is shown in Fig.~\ref{xsecconv}. The first 5/2$^{+}$ resonance is the one that best reproduces  the experimental data. However, the second resonance gives a similar angular distribution and even the sum of both would be consistent with the data. As shown in Ref.~\cite{Mor12a}, the magnitude and shape of the resonant break-up is sensitive to the weights of the different configurations of each state. Unfortunately, in this case, both resonances are mainly based in the 2$^+$ core excited state and, therefore, there is not a clear difference between both choices.  Furthermore, in this case the population of both resonances was found to be almost exclusively due to the core excitation mechanism. 
To illustrate this effect, we include in Fig.~\ref{xsecconv} a standard inert-core DWBA calculation where the ground state and the $5/2^+$ resonant state are represented by pure $s_{1/2}$ and $d_{5/2}$ single-particle configurations orbiting an inert $^{18}$C core, respectively.
%\sout{ of $^{19}$C is considered to be a neutron in a pure $s_{1/2}$ orbit, and for the resonance in a pure $d_{5/2}$ orbit, around an inert $^{18}$C core in its ground state.}
 The result of this calculation is given by the dotted line in  Fig.~\ref{xsecconv}. It is clearly seen that the resulting angular distribution significantly underestimates the magnitude of the data, and fails to reproduce the shape too. The same conclusion was achieved in Ref.~\cite{Cre11b} where a AGS-Faddeev calculation, using a more realistic $p-n$ interaction (CD-Bonn), but ignoring core excitations, was also found to provide too small a break-up cross section. This result clearly shows that  the observed resonant peak is not consistent with a simple $2s_{1/2}\rightarrow 1d_{5/2}$ transition and evidences the dominance of the core excitation mechanism in the present case, resulting from the large $^{18}$C(2$^+$) component in both resonances (c.f.\ Table~\ref{Tab:sf19c}). The DCE mechanism is much larger than that found in the $^{11}$Be+$p$ case, in which the valence and core excitations have been found to be of similar magnitude. 

% within the same Fig.~\ref{xsecconv} looking at the contribution of the single particle excitation to the total cross section for the first resonance (dotted line). The core excitation dominates in this case due to the large component coming from the 2$^+$ excited state of $^{18}$C in both resonances (c.f.\ Table~\ref{Tab:sf19c}).

\begin{table}
\caption{\label{Tab:sf19c} Weights of the different configurations for the ground state and the two 5/2$^{+}$ resonances in  $^{19}$C, according to the semi-microscopic particle-plus-core model described in the text  \cite{Lay14}.}
\begin{center}
\begin{tabular}{ccccc}
\hline
 & $| 0^+ \otimes (\ell s)j \rangle $ &    $ | 2^+ \otimes s_{1/2} \rangle $  &   $ | 2^+ \otimes d_{3/2} \rangle $ & $ | 2^+ \otimes d_{5/2} \rangle $    \\
\hline  
\hline
$1/2^{+}_{1}$     & 0.529    &    --   &  0.035  &   0.436    \\
%               &  PRM(1)   & 0.517    &    --   &  0.081  &   0.402    \\  
%               &  PRM(2)   & 0.505    &    --   &  0.033  &   0.462    \\  
%               &  WBP      & 0.580    &    --   &  0.085  &   0.470    \\  

%\hline
%$3/2^{+}_{1}$  &   P-AMD   & 0.028    &  0.386  &  0.121  &   0.464   \\  
%               &   PRM(1)  & 0.043    &  0.348  &  0.150  &   0.459    \\  
%               &   PRM(2)  & 0.023    &  0.371  &  0.106  &   0.500    \\  
%               &   WBP     & 0.026    &  0.494  &  0.001  &   0.076   \\

\hline
$5/2^{+}_{1}$    & 0.276    &  0.721  &  0.000  &   0.003  \\ 
%               &   PRM(1)  & 0.285    &  0.716  &  0.000  &   0.003    \\  
%               &   PRM(2)  & 0.278    &  0.719  &  0.000  &   0.003    \\  
%               &   WBP     & 0.383    &  0.015  &  0.000  &   0.751       \\ 

\hline
$5/2^{+}_{2}$   & 0.200    &  0.142  &  0.002  &   0.657   \\  
%               &   PRM(1)  & 0.217    &  0.178  &  0.004  &   0.602    \\  
%               &   PRM(2)  & 0.207    &  0.100  &  0.002  &   0.690    \\  
%               &   WBP     & 0.035    &  0.609  &  0.009  &   0.291  \\

\hline
\end{tabular}
\end{center}
\end{table}

%{\bf AMM: Yo esto lo quitaria, porque creo que añade poco a la discusión:}
%
%{\it 
%XXX OPTIONAL XXX
%
%In addition, it is possible to compare the semi-microscopic model for $^{19}$C with the adjusted particle-rotor model. In Fig.~XXX we show that both models give almost exactly the same results. This is connected with the fact that essentially same spectroscopic factors are obtained by adjusting the phenomenological particle-rotor interaction to the semi-microscopic folding potential. As we have already mentioned, this resonant break-up at intermediate energies is particularly sensitive to this observable.
%
%XXX OPTIONAL XXX
%}

%XXX Citing \cite{Cre11b} XXX

%--------------------- CONCLUSIONS -------------------
\textit{Conclusions.} 
%---------------------------------------------------
We have investigated the role of core excitations in the resonant break-up of $^{19}$C on a proton target. For that, we have considered a two-body model for $^{19}$C and performed XCDCC and XDWBA calculations that include the possibility of core ($^{18}$C) excitations  in the structure of the projectile as well as in the reaction dynamics.

 We have compared our results with the experimental data measured by  Satou and collaborators~\cite{Sat08} for this reaction,  at an incident energy of 70~MeV/u, corresponding to the  angular distribution for a resonant state in $^{19}$C, which was identified with the second $5/2^+$ state predicted by sd shell-model calculations.

 Our structure calculations, based on a particle-plus-core model of $^{19}$C,  predict two  $5/2^+$ low-lying resonances, but none of them at the energy of the peak observed in ~\cite{Sat08}. Furthermore, the corresponding angular distributions are both compatible with the shape and magnitude of the experimental one, thus precluding an unambiguous  identification of the experimental peak with one or another. This result is understood as a consequence of the similar structure for the two resonances. Both resonances are mainly based on the first 2$^{+}$ state of the core. Therefore, it is clearly seen in the present analysis that the dynamic excitation of the core is the main responsible for the peak observed in the break-up with protons. Moreover, we have shown that the pure valence excitation mechanism, assuming a $2s_{1/2}\rightarrow 1d_{5/2}$ single-particle transition, gives a negligible contribution here. This is the first case where we have identified that the core excitation mechanism dominates overwhelmingly.

The present results are in contrast with the naive picture of halo nuclei where the weakly-bound neutron is completely decoupled from the rest of nucleons inside the core, which could be considered as a frozen object. We had previously found cases where single-particle excitations of the valence particle and dynamic excitations of the core compete on equal footing, leading to an interesting interplay of both processes~\cite{Mor12a}. However, the dynamic excitation of the core in $^{19}$C is so strong that it is the core the one that plays the main role in the break-up reaction of a halo nucleus.

As a final remark, we would like to insist on the importance of the effects of core excitations in reactions with halo nuclei. The cores of the new and heavier halo candidates, like $^{31}$Ne and $^{37}$Mg, will present more and more complex structures since they will be more exotic. This will make the analysis of the forthcoming experiments more involved. Taking into account possible core excitation effects will be mandatory for a better understanding and a correct analysis of the experimental data.

%-----------------------------
 \textit{Acknowledgments.} 
%----------------------------
This work has been partially supported by the Spanish Ministerio de Ciencia e Innovaci\'on and FEDER funds under projects
 FIS2014-53448-C2-1-P,  FIS2013-41994-P,  by the Spanish Consolider-Ingenio 2010 Programme CPAN
(CSD2007-00042), by Junta de  Andaluc\'ia (FQM160,
 P11-FQM-7632), and by the Funda\c{c}\~{a}o para a Ci\^{e}ncia e a Tecnologia (FCT) grants SFRH/BPD/78606/2011 and PTDC/FIS-NUC/2240/2014. The research leading to these results has also received funding from the European Commission, Seventh Framework Programme (FP7/2007-2013) under Grant Agreement No. 600376. JAL is a Marie Curie-Piscopia fellow at the University of Padova. RCJ is supported by the UK STFC through Grant No. ST/F012012/1.

\bibliography{c19p_v7}
%\bibliography{pamd}

\end{document}